\documentclass[aps,prb,twocolumn,groupedaddress,showpacs]{revtex4}
\usepackage{epsfig}
\usepackage{epsf}
\usepackage{amssymb, amsmath}
\usepackage{graphicx}
\usepackage{dcolumn}
\usepackage{bm}
\usepackage{appendix}
\newcommand{\be}{\begin{equation}}
\newcommand{\ee}{\end{equation}}
\newcommand{\bea}{\begin{eqnarray}}
\newcommand{\eea}{\end{eqnarray}}

\newcommand{\ua}{\uparrow}
\newcommand{\da}{\downarrow}

\newcommand{\vk}{{\boldsymbol k}}

\usepackage{color}
\usepackage{ulem}
\begin{document}
\title{Hall conductivity as bulk signature of topological transitions in superconductors}
\author{Pedro D. Sacramento$^{1}$, Miguel A. N. Ara\'ujo$^{1,2}$, Eduardo V. Castro$^{1}$, 
} 
\affiliation{$^1$ CFIF, Instituto Superior 
T\'ecnico, TU Lisbon, Av. Rovisco Pais, 1049-001 Lisboa, Portugal}
\affiliation{$^2$ Departamento de F\'{\i}sica,  Universidade de \'Evora, P-7000-671, \'Evora, Portugal}

\begin{abstract}
Topological superconductors may undergo transitions between
phases with different topological numbers which, like the case of topological insulators, 
are related to the presence of gapless (Majorana) edge states. 
In $\mathbb{Z}$ topological insulators  the charge Hall conductivity is quantized,
being proportional to the number of gapless states running at the edge.
In a superconductor, however, charge is not conserved
 and, therefore, $\sigma_{xy}$ is not quantized, even
in the case of a $\mathbb{Z}$ topological superconductor.
Here it is shown that while the 
 $\sigma_{xy}$ evolves continuously between different topological phases of a 
$\mathbb{Z}$ topological superconductor,
its derivatives  display sharp features signaling the topological transitions.
We consider in detail the case of a triplet superconductor with $p$-wave 
symmetry in the presence of Rashba spin-orbit (SO)
coupling and externally applied Zeeman spin splitting. 
Generalization to the cases
where the pairing vector is not aligned with that of the SO coupling is given.
We generalize also to the cases where
the normal system is already topologically non-trivial.  

\end{abstract}

\maketitle

\section{Introduction}

In two dimensions, 
$\mathbb{Z}$~topological insulators exhibit a charge Hall conductivity that 
is quantized and
proportional to the Chern number of the 
occupied bands \cite{TKNN,Hasan,Zhang}. 
Such nontrivial topological phases are also characterized
by the presence of gapless edge modes \cite{Halperin,Hatsugai} that can be detected by transport measurements or tunneling. 
In a topologically non-trivial superconductor 
one does not expect that the charge Hall conductivity may be quantized,
however, as charge is not conserved due  to the breaking of $U(1)$ symmetry. 
In a singlet superconductor spin is conserved and
there is still a possibility that the spin Hall
conductivity is quantized, 
as previously shown for a $d$-wave superconductor in the vortex state \cite{ZlatkoHall}.
For a triplet superconductor even this quantization is absent.
The thermal Hall conductivity has recently been shown to be quantized for 
topological superconductors with broken time reversal symmetry (TRS) \cite{Sumiyoshi}.

Generically speaking, 
the charge Hall resistance  may be written as the sum of two contributions, one proportional to the
magnetic field and an anomalous contribution as
$\rho_{xy}=R_0 H_z + \rho_{xy}^{AH}$  (considering
$z$ as the perpendicular direction to the plane where the charges move). 
The term $\rho_{xy}^{AH}$ 
is the anomalous Hall effect \cite{Nagaosa}
and has different origins. One of these origins  is intrinsic and is the
result of an anomalous velocity \cite{Karplus} that is the result of a non-zero Berry curvature
\cite{Xiao}, $\boldsymbol{\Omega}_n(k)$.
The velocity of a charged particle in a given energy band $n$ in the presence of an electric field, 
$\boldsymbol{E}$, can be written as 
$\boldsymbol{v}_n(k)=\hbar^{-1} \partial \epsilon_n(k)/\partial k
-(e/\hbar) \boldsymbol{E} \times \boldsymbol{\Omega}_n(k)$.
The last term gives a contribution to the  
velocity that is transverse to the direction of the electric field
and therefore 
contributes to the Hall conductivity. Two other mechanisms that lead to an anomalous velocity
are due to scattering from impurities in a system where the spin-orbit has to be taken into account
such as the skew scattering mechanism \cite{Smit} and the side jump \cite{Berger}.

Historically, the anomalous Hall resistivity was studied in detail in systems with a
finite magnetization $M_z$, where $\rho_{xy}^{AH} = R_sM_z$.\cite{Nagaosa}
Anomalous properties in superconductors with magnetization have been studied 
before in particular,
the presence of magnetoelectric effects \cite{edelstein}, the generation of a charge Hall effect due to a 
spin current \cite{maekawa}
or a spin Hall effect due to a charge current \cite{bernevig}. 
The anomalous Hall effect has been studied \cite{us}
in superconductors with spin-orbit coupling. The presence of a magnetic impurity is enough to induce a
non-vanishing Hall conductivity \cite{us}. The magnetic impurity interacts with the superconductor as
a local Zeeman term and orbital effects such as the presence of vortices are neglected. The local magnetization
may also be the result of some proximity effect for instance with a magnetic dot.
A dense magnetic impurity distribution leads to the destruction of superconductivity if the pairing
is a spin singlet but magnetization and superconductivity may coexist if the pairing is of triplet
origin such as in a p-wave superconductor. In this case the Hall conductivity
was shown to be non-vanishing if the spin-orbit coupling is present \cite{us}. 

In this work we will
be concerned with the effect of the intrinsic contribution to the charge Hall conductivity
in $\mathbb{Z}$~topological superconductors. Superconductivity
with non-trivial topology may be obtained in different ways.\cite{Zhang} It can be due to
the pairing symmetry, as is the case of $p$-wave superconductors.\cite{ReadGreen}
In semiconductors with Rashba spin-orbit coupling it arises when $s$-wave superconductivity 
is induced
and a Zeeman (time-reversal breaking) term is added.\cite{SatoPRL09, SauPRL10} An interesting proposal
is that of systems where the normal phase is already topologically non-trivial, in which case a 
topological superconductor can
be obtained if $s$-wave superconductivity is induced by proximity effect.\cite{Fu,QHZ10}
Here we consider a Rashba-type non-centrosymmetric superconductor with admixture of $s$-wave
and $p$-wave pairing and TRS  breaking Zeeman splitting, which has been
recently proposed in Ref.~\onlinecite{Sato}. We generalize to the cases where the pairing vector 
is not coincident with that of the spin-orbit coupling, and also to the cases where the normal
phase is topologically non-trivial. We thus take on equal footing the three possible ways of
obtaining topological superconductivity mentioned above.

The main results of this paper may be summarized as follows. Whenever a $\mathbb{Z}$~topological
superconductor is realized and the first Chern number fully characterizes the topological phase,
we find that the behavior of the Hall conductivity and its derivatives with respect to the parameters
that drive the topological phase transition, specially the second derivative, provide an alternative
way to identify topological transitions in superconductors. This approach proves extremely 
useful when the pairing vector is not aligned with the spin-orbit, in which case  we have found
a less obvious relation between the Chern number and the number of crossings of edge state
bands with the Fermi level. A careful topological analysis of this case is given.

The paper is organized as follows. 
In  Section II we introduce the expression for the Chern number that will be used throughout
this work. The model Hamiltonian is presented in Section III and in  Section IV the
results for the Hall conductivity and Chern number are presented. Section V is devoted
to the analysis of the edge states and their correspondence to the topological indices.
In Section VI we present results for a model where the non-superconducting band structure is already
nontrivial. Our conclusions are presented in Section VII.

\section{Characterization of topological phases}

The Berry curvature tensor
for a band with Bloch wavefunctions $u_n(\vk)$ 
can be calculated as
\be
\boldsymbol{\Omega}_n(\vk) = \langle \nabla_\vk u_n(\vk) | \times | \nabla_\vk u_n(\vk) \rangle\,,
\ee
where 
$\boldsymbol k=(k_x,k_y)$ denotes the momentum vector. 
 The contribution  from the $n$-th band  
to the Hall conductivity of a normal system  
may be written in terms of the Berry curvature as\cite{TKNN}:
\be
\sigma_{xy}^{(n)} = \frac{e^2}{\hbar} \int_{BZ} \frac{d^2 k}{(2\pi)^2} \Omega_n^{x,y}(\vk) n_F(\epsilon_n(\vk))\,,
\ee
where $n_F$ is the Fermi function.
If the chemical potential lies within a gap the integral over the occupied states runs over 
the entire Brillouin zone.
The charge Hall conductivity can then be written as
\be
\sigma_{xy} = C \frac{e^2}{h}\,,
\ee
where $C$ is the sum of the Chern numbers of the occupied bands.
The Berry curvature may also be obtained as a sum over states analogous to the Kubo
formula for the conductivity, and reads:
\be
\Omega_n^{\mu,\nu}=i \sum_{n' \neq n} \frac{\langle n| \frac{\partial H}{\partial k_{\mu}}|n'\rangle \langle
n' | \frac{\partial H}{\partial k_{\nu}} |n\rangle - \mu \leftrightarrow \nu}{(E_n-E_{n'})^2}\,.
\ee
In  the case of a superconductor, the states $|n\rangle$ are the eigenstates of the Bogoliubov-de Gennes equations. At the gapless
points the denominator vanishes and the integral over the Brillouin zone may have large numerical errors. 
It is then more convenient to calculate the Chern number by computing the
flux of the Berry curvature over plaquetes in the Brillouin zone \cite{Fukui}.
Discretizing the Brillouin zone as $k_\mu = 2\pi j/N$, with $j=1,...,N$, and $\mu=x,y$,    
a new variable, $U_{\mu}(\vk)$, for the link $\delta k_\mu$ (with $|\delta k_\mu|= 2\pi/N$) 
oriented along the  $\mu$ direction from the point $\vk$ may be defined as
\be
U_{\mu}(\vk) = \frac{\langle n(\vk)|n(\vk +\delta k_\mu)\rangle}{|\langle n(\vk)|n(\vk +\delta k_\mu)\rangle |}\,, 
\ee
and the lattice field strength may be defined as
\be
F_{xy}(\vk) = \ln \left( U_x(\vk) U_y(\vk +\delta k_x) U_x(\vk +\delta k_y)^{-1} U_y(\vk )^{-1} \right)\,.
\ee
$F_{xy}(\vk)$  is restricted to the interval     
$-\pi < -i F_{xy}(\vk) \leq \pi$ and the gauge invariant expression for the Chern number is
\be
C_n=\frac{1}{2\pi i} \sum_\vk F_{xy}(\vk)\,.
\label{fukui}
\ee
The calculations of the Chern number of each band $n$ are performed in this way in this work.

The calculation of the Chern number is simple for a 2$\times$2 
Hamiltonian  matrix $\hat H$ once the latter is written in the form:  
\begin{equation}
\hat H(\boldsymbol h) = \boldsymbol h( \boldsymbol k) \cdot  \boldsymbol \tau
  + h_0(\boldsymbol k) \tau_0 \,,
\label{H}
\end{equation}
where 
$ \boldsymbol h=(h_x,h_y,h_z)$,
$\tau$ are Pauli matrices and $\tau_0$ is
the identity. The
Chern number for the bands in  Hamiltonian Eq.~(\ref{H})  is independent of the choice for
$h_0(\boldsymbol k)$, as computed from the
usual expression
\begin{equation}
C=\frac{1}{4\pi}\int dk_x\ dk_y \  \frac{\partial \hat{\boldsymbol h}}{\partial k_x}
\times  \frac{\partial \hat{\boldsymbol h}}{\partial k_y} \cdot \hat{\boldsymbol h}\,,
\label{chern}
\end{equation}
$\hat{\boldsymbol h}=\boldsymbol h/|\boldsymbol h|$.
The topological nature of bands
may be understood as the result of the covering of the unit sphere defined by the vector
$\hat{\boldsymbol h}$. 

The system's symmetry properties depend on whether
 the Pauli matrices in equation (\ref{H}) represent a pseudospin 
({\it e.g.}, a sublattice)  
or the physical  spin. In the first case
TRS requires $h_{x(z)}$ to be  an even function
of $\boldsymbol k$ and $h_y$ to be  odd.
Otherwise, all components have to be odd.
In order to have nonzero $C$, TRS must be broken. The operation
of spatial inversion does not change the Chern number $C$.

On general grounds, non-trivial topological order for non-interacting Hamiltonians can be 
related with the presence or absence of three discrete symmetries: time-reversal, particle-hole,
and chiral symmetry.\cite{Ludwig,LudwigAIP,LudwigNJP} For Bogoliubov-de Gennes systems, where particle-hole symmetry
is always present, preserving or not TRS is determinant to the nature of possible 
topological phases in two dimensions. The non-centrosymmetric superconductor we consider here
is time-reversal invariant if the Zeeman term is absent and the pairing is unitary.
The system then belongs to the symmetry class DIII where the topological invariant is
a $\mathbb{Z}_2$ index, and it is said to realize a $\mathbb{Z}_2$~topological 
superconductor. If the pairing is non-unitary or the Zeeman term is finite TRS is broken and the system belongs
to the symmetry class D (the TRS operator $\mathcal{T}$ is such that $\mathcal{T}^2 = -1 $).
The topological invariant that characterizes this phase is the first Chern number $C$,
and the system is said to be a $\mathbb{Z}$~topological superconductor.

\section{Topological superconductor}
\label{sec:TS}

We consider a triplet superconductor with $p$-wave symmetry in the presence of Rashba spin-orbit coupling
and magnetization, {\it e.g.}, due to a  time-reversal breaking Zeeman term.
Due to the non-centrosymmetric nature of the system, parity is broken and, in general,
the pairing symmetry is not fixed, and an admixture of singlet pairing is allowed \cite{Gorkov}. Therefore, we
also consider a contribution from $s$-wave pairing. 
This model was studied in Refs. \cite{Sato,us}. 
We write the Hamiltonian as
\begin{eqnarray}
\hat H = \frac 1 2\sum_\vk  \left( {\boldsymbol c}_{\vk}^\dagger ,{\boldsymbol c}_{-\vk}   \right)
\left(\begin{array}{cc}
\hat H_0(\vk) & \hat \Delta(\vk) \\
\hat \Delta^{\dagger}(\vk) & -\hat H_0^T(-\vk) \end{array}\right) 
\left( \begin{array}{c}
 {\boldsymbol c}_{\vk} \\  {\boldsymbol c}_{-\vk}^\dagger  \end{array}
\right)
\label{bdg1}
\end{eqnarray}
where $\left( {\boldsymbol c}_{\vk}^{\dagger}, {\boldsymbol c}_{-\vk} \right) = 
\left( c_{\vk\ua}^{\dagger}, c_{\vk\da}^\dagger ,c_{-\vk\ua}, c_{-\vk\da}   \right)$ 
and
\begin{equation}
\hat H_0=\epsilon_\vk\sigma_0 -M_z\sigma_z + \hat H_R\,.
\end{equation}
Here, $\epsilon_{\boldsymbol{k}}=-2 t (\cos k_x + \cos k_y )-\epsilon_F$
is the kinetic part, where $t$ denotes the hopping parameter set in
the following as the energy scale, $t=1$, $\epsilon_F$ is the
chemical potential,
$\boldsymbol{k}$ is a wave vector in the $xy$ plane, and we have taken
the lattice constant to be unity, $a=1$. Furthermore, $M_z$ 
is the Zeeman splitting term responsible for the magnetization, 
in energy units, along the $z$ direction.
Finally, the Rashba spin-orbit
term is written as 
\begin{equation}
\hat H_R = \boldsymbol{s} \cdot \boldsymbol{\sigma} = \alpha
\left( \sin k_y \sigma_x - \sin k_x \sigma_y \right)\,,
\end{equation}
 where
$\alpha$ is measured in the energy units 
 and $\boldsymbol{s} =\alpha(\sin k_y,-\sin k_x, 0)$.
The matrices $\sigma_x,\sigma_y,\sigma_z$ are
the Pauli matrices acting on the spin sector, and $\sigma_0$ is the
$2\times2$ identity.

The pairing matrix reads
\begin{equation}
\hat \Delta = i\left( {\boldsymbol d}\cdot {\boldsymbol\sigma} + \Delta_s \right) \sigma_y =
 \left(\begin{array}{cc}
-d_x+i d_y & d_z + \Delta_s \\
d_z -\Delta_s & d_x +i d_y
\end{array}\right)\,.
\end{equation}
The vector $\boldsymbol{d}=(d_x,d_y,d_z)$ is the vector representation of the
$p$-wave
superconducting pairing and is an odd function of $\boldsymbol k$. 
Because of Fermi statistics, the  pairing matrix satisfies
$\hat \Delta(\vk) = -\hat \Delta^T (-\vk) $.
The triplet pairing term is invariant under a spin rotation about the $\boldsymbol{\hat d}$ direction.
We note that both the superconducting order parameter and the
magnetization may be due to intrinsic order or to some proximity effect due to
neighboring superconductors or ferromagnets.
The pairing matrix for a p-wave superconductor generally satisfies
\be
\hat \Delta \hat \Delta^{\dagger} = |\boldsymbol d|^2 \sigma_0 + \boldsymbol q \cdot \boldsymbol{\sigma}\,,
\ee
where $\boldsymbol q=i \boldsymbol d \times \boldsymbol d^*$. 
If the vector $\boldsymbol q$ vanishes the pairing is called unitary (s-wave
pairing is always unitary). Otherwise it is called non-unitary\cite{Sigrist}
and breaks TRS, 
originating a spontaneous magnetization in the system due to the symmetry of the pairing, as in $^3He$.

We will consider both unitary and non-unitary pairings. In the case of unitary
pairing we consider two examples. One of them respects to a situation where the
spin-orbit coupling is such that the pairing is aligned \cite{Sigrist2} along the 
spin-orbit vector $\boldsymbol{s}$. This is a situation expected if the spin-orbit
is strong since it is energetically favorable, and we will denote it by strong coupling case. In the other case we will
relax this restriction and allow that the two vectors are not aligned. This case
we will denote by weak spin-orbit coupling.

The energy eigenvalues and eigenfunction may be obtained solving the Bogoliubov-de Gennes equations
\be
\label{bdg2}
\left(\begin{array}{cc}
\hat H_0(\vk) & \hat \Delta(\vk) \\
\hat \Delta^{\dagger}(\vk) & -\hat H_0^T(-\vk) \end{array}\right)
\left(\begin{array}{c}
u_n\\
v_n
\end{array}\right) 
= \epsilon_{k,n} 
\left(\begin{array}{c}
u_n\\
v_n
\end{array}\right).
\ee
The 4-component spinor can be written as \cite{errata}
\be
\left(\begin{array}{c}
u_n\\
v_n
\end{array}\right)=
\left(\begin{array}{c}
u_n(\boldsymbol{k},\uparrow) \\
u_n(\boldsymbol{k},\downarrow) \\
-v_n(-\boldsymbol{k},\uparrow) \\
v_n(-\boldsymbol{k},\downarrow) \\
\end{array}\right)
\ee
The energy eigenvalues of the Hamiltonian \eqref{bdg1} can be written (for
$\Delta_s=0$ and $d_z=0$) as 
\be 
\label{bands} \epsilon_{\boldsymbol{k},\alpha_1,\alpha_2}
= \alpha_1 \sqrt{z_1 +\alpha_2 2 \sqrt{z_2}}, 
\ee  
where 
\bea
z_1 &=& \boldsymbol{d}\cdot \boldsymbol{d}^* + \boldsymbol{s}\cdot \boldsymbol{s} + \epsilon_{\boldsymbol{k}}^2 
+ M_z^2 \,,\nonumber \\
z_2 &=& \left| (\boldsymbol{d}\times \boldsymbol{s})_z \right|^2 -
i \epsilon_{\boldsymbol{k}} M_z (\boldsymbol{d}\times \boldsymbol{d}^*)_z + \nonumber\\
&&\frac{1}{4}\left[ (\boldsymbol{d}\cdot \boldsymbol{d}^*)^2 - \left| \boldsymbol{d}\cdot \boldsymbol{d} \right|^2  \right] +
\epsilon_{\boldsymbol{k}}^2(\boldsymbol{s}\cdot \boldsymbol{s} + M_z^2 ), 
\eea 
and $\alpha_1,\alpha_2=\pm$.
The gap between the lowest bands closes at the $\boldsymbol k$ points satisfying the condition 
$z_1=2 \sqrt{z_2}$.

In the superconducting phase the system is generally gapped. A possible change of topology 
occurs when the gap closes.
Considering the strong spin orbit case for which
the $\boldsymbol d$ and $\boldsymbol s$ are collinear\cite{Sato},  
we may write 
$\boldsymbol{d}=(d/\alpha) \boldsymbol{s}$.
Taking  also the   $s$-wave pairing into account, 
the  gapless points satisfy
\bea
\epsilon_{\boldsymbol{k}}^2+\Delta_s^2 &=& M_z^2 + \left(1+ \frac{d^2}{\alpha^2} \right) \boldsymbol{s}^2, \nonumber \\
\epsilon_{\boldsymbol{k}} \frac{d}{\alpha} \boldsymbol{s} &=& \Delta_s \boldsymbol{s}\,.
\label{gapless}
\eea

\section{Chern numbers and Hall conductivity}
\label{sec:Csigma}

\begin{figure}
\begin{centering}
\includegraphics[width=0.8\columnwidth]{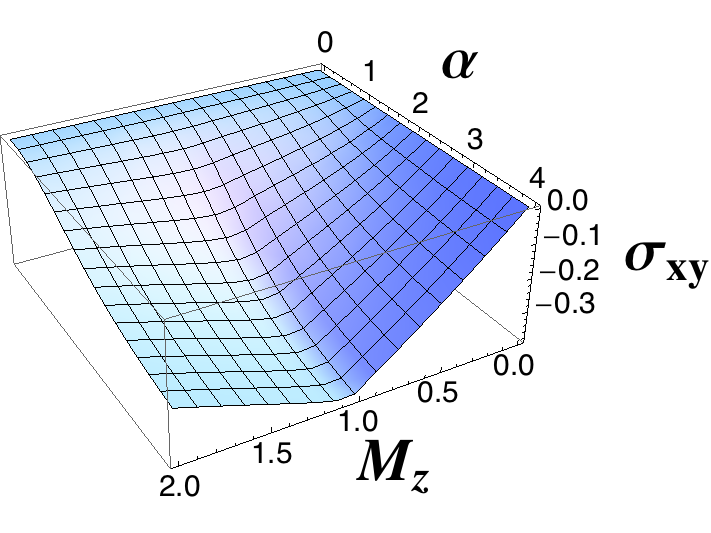}
\includegraphics[width=0.8\columnwidth]{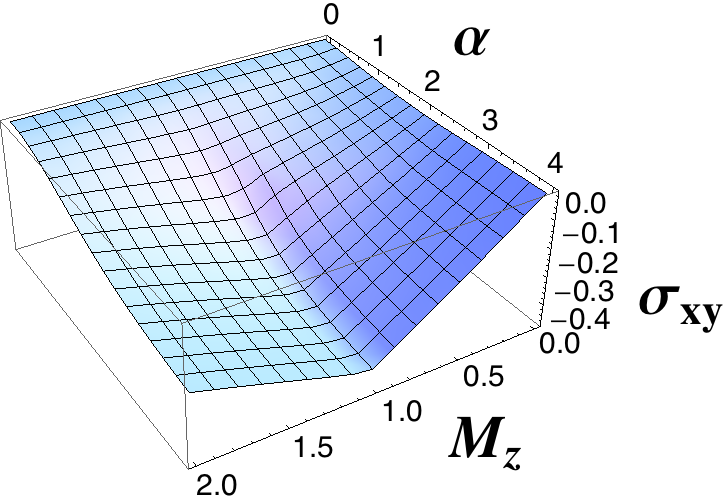}
\par\end{centering}
\caption{\label{fig:hallsold}(color online). Hall conductivity in a unitary case with $d_x=d \sin k_y, d_y=d \sin k_x, d_z=0$
(top panel)
and a non-unitary case with $d_x=-d/2 \sin k_x=-i d_y, d_z=0$ (bottom panel) with $d=1, \epsilon_F=-1$.
This Figure corrects a  previously obtained result \cite{us}.}
\end{figure} 

\subsection{Weak spin-orbit coupling: Unitary and non-unitary pairings}

In Fig. \ref{fig:hallsold} we present the results for the Hall conductivity,
as calculated from the Kubo formula, given in Eq. (10) of Ref. \cite{us}.
We consider both the unitary and the 
non-unitary cases, relaxing the restriction that 
$\boldsymbol{d} \parallel \boldsymbol{s}$ which was assumed in Ref.~\onlinecite{Sato}. 
The two cases are chosen as $d_x=d \sin k_y, d_y=d \sin k_x, d_z=0$ and
$d_x=-d/2 \sin k_x=-i d_y, d_z=0$, respectively. 
The  Hall conductivity in the superconducting unitary
phase is similar to that of the normal phase (not shown). 
Both in the normal phase and in the unitary case a magnetization
is required in order to have a non-vanishing Hall conductivity. 
However, in the non-unitary case,  the magnetization
induced by the vector $\boldsymbol q$ 
produces a finite
Hall conductivity even if the explicit Zeeman term is absent.
In all cases the spin-orbit coupling is necessary for a non-vanishing Hall conductivity.

\begin{figure}
\begin{centering}
\includegraphics[width=0.8\columnwidth]{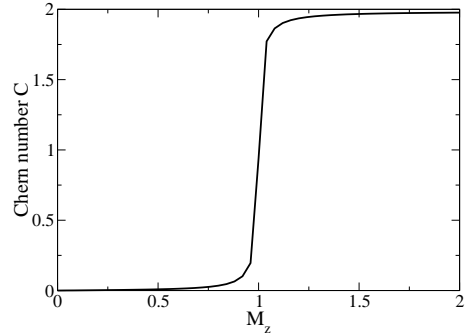}
\par\end{centering}
\caption{\label{fig:Chernold}
Chern number for the unitary case of Fig. \ref{fig:hallsold} as a function of magnetization for $d=1, \epsilon_F=-1$ and $\alpha=1$.
}
\end{figure}

In all three cases (normal, unitary, and non-unitary) the Hall conductivity has a clear 
minimum when the magnetization is of the order of the
chemical potential.  
At this point the spectrum is gapless and, 
as shown in Fig. \ref{fig:Chernold} for the unitary case, the Chern number of the occupied bands
changes. 
The topological transition does not depend on $\alpha$ and the results in Fig. \ref{fig:Chernold} are shown for
the particular case of $\alpha=1$. 
It turns out that for the non-unitary case depicted in Fig. \ref{fig:hallsold} the spectrum is always
gapless due to the lack of dependence of the pairing function on $k_y$. 
Therefore, the expression for the Chern number suffers from numerical instability.
Considering a non-unitary pairing of the form $d_x=d \sin k_y, d_y=i d \sin k_x, d_z=0$ 
the Hall conductivity is similar to that obtained in Fig. \ref{fig:hallsold} and the change in the
Chern number is also similar to the one shown in Fig. \ref{fig:Chernold} for the unitary case.

In the normal phase the system is topologically
trivial and the Chern number is zero throughout the space of parameters of the chemical potential and the
magnetization. 

\subsection{Strong spin-orbit coupling}

For strong spin-orbit coupling it has been shown that it is more favorable that the pairing
vector $\boldsymbol d$ aligns with the spin-orbit coupling:
\be
\boldsymbol d(\boldsymbol k) = d (\sin k_y, -\sin k_x)\,.
\ee
As a consequence, the critical temperature associated with this type of pairing is higher \cite{Sigrist2}.
There is a rich sequence of topological transitions 
as a function of the chemical potential,
spin-orbit coupling and magnetization \cite{Sato}. 
In general this  problem  involves solving for the eigenvalues
of the $4\times 4$   matrix in Eq.~(\ref{bdg1}). The calculation of the Chern number of each band is performed
using the eigenfunctions of this $4\times 4$ matrix in Eq.~\eqref{fukui}.
Since the gap must close at the 
 topological transitions, 
the location of these transitions may be determined looking at the gapless $\boldsymbol k$
points \cite{Sato} satisfying Eq.~(\ref{gapless}). 
The location of the transitions
and the associated gapless points in the spectrum have been obtained before \cite{Sato}. It turns out that
in each topological phase, the Hamiltonian can be continuously deformed in such a way that
($\alpha$,  $\Delta_s \rightarrow 0$) without closing the gap.
The problem then simplifies
since the original $4\times 4$ Hamiltonian has been deformed 
to two $2\times 2$ matrices and the Chern number may be
calculated as in Eq.~\eqref{chern}. 

\begin{figure}
\begin{centering}
\includegraphics[width=0.8\columnwidth]{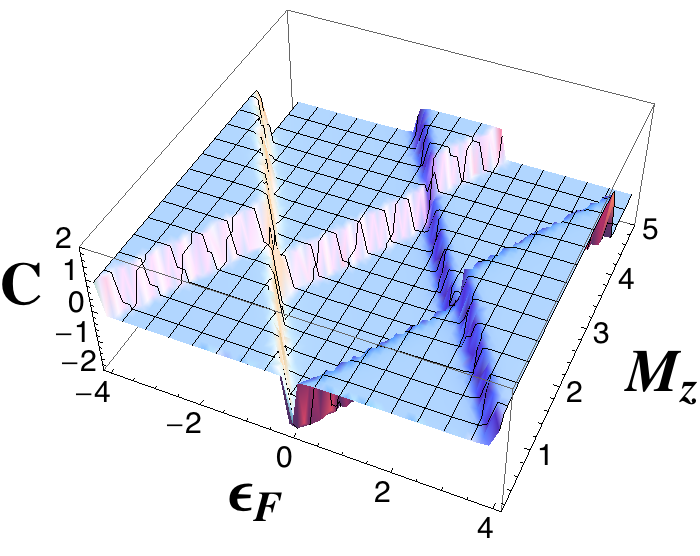}
\includegraphics[width=0.8\columnwidth]{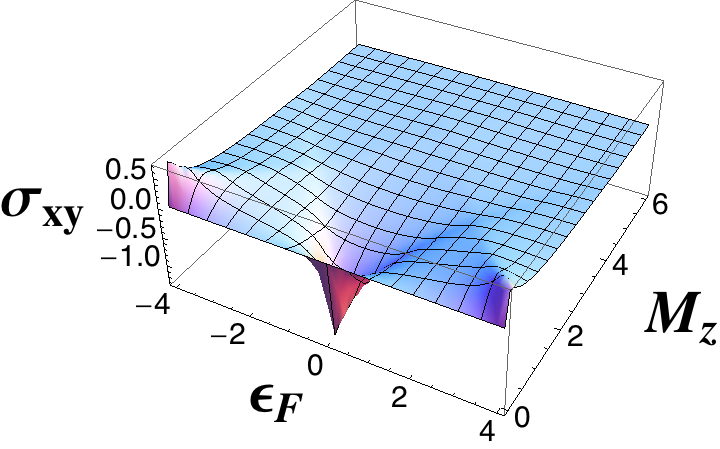}
\par\end{centering}
\caption{\label{fig:hallstrong}(color online). Chern number as a function of chemical potential and magnetization and 
Hall conductivity for the case of strong spin-orbit coupling. The parameters used are $d=0.6, \Delta_s=0.1, \alpha=0.6$.
}
\end{figure} 

In Fig.~\ref{fig:hallstrong}
we show the results for the Chern number of the occupied bands as a function of the chemical potential and
magnetization. There are various transitions that correspond to the closing and opening of gaps in the
spectrum. In the same figure we show the results for the Hall conductivity for the same region of parameters.
Even though the Hall conductivity is a continuous, smooth function there are clearly local maxima and minima
that can be associated to the points where a topological transition occurs. 

\begin{figure}
\begin{centering}
\includegraphics[width=0.9\columnwidth]{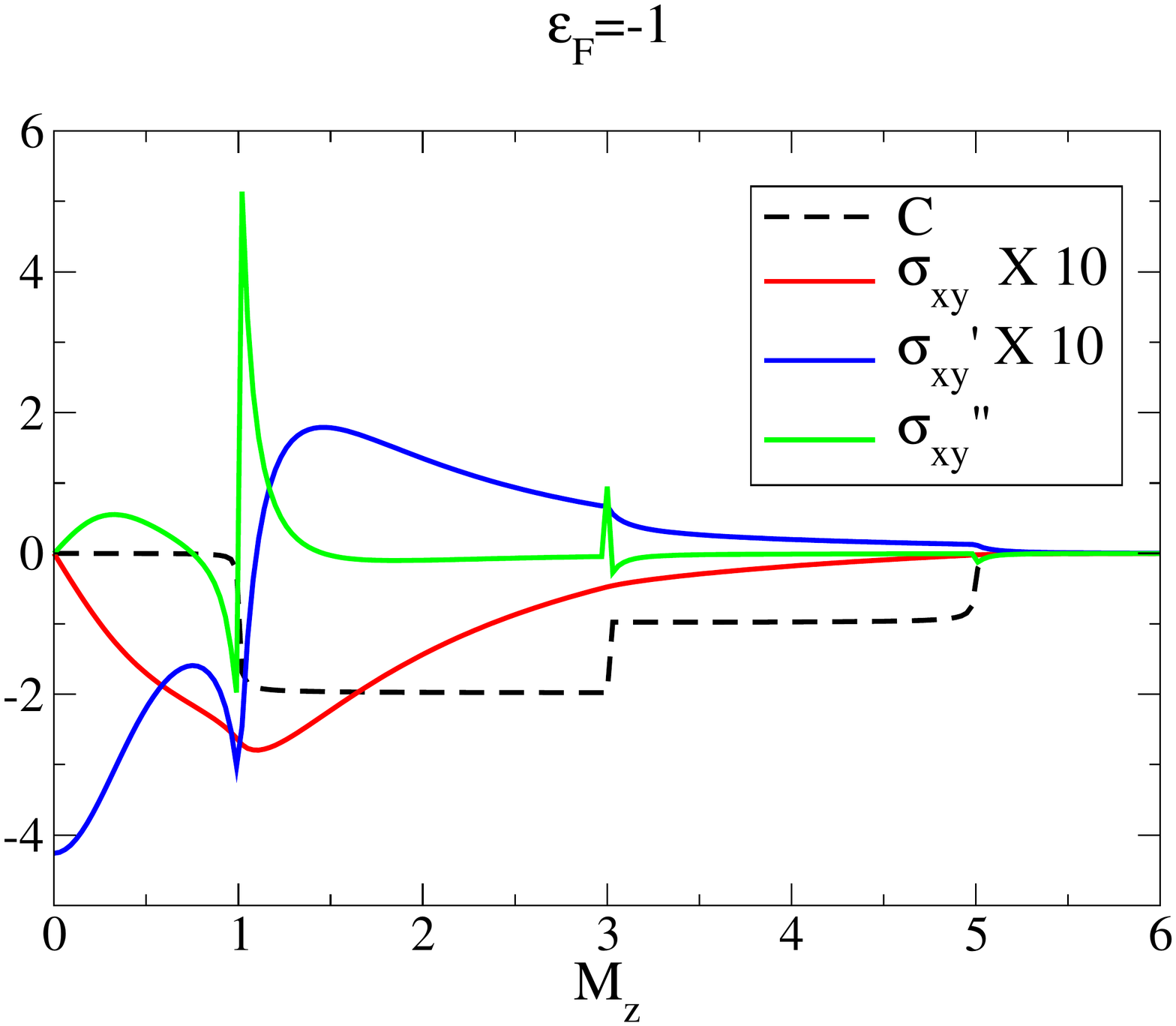}
\includegraphics[width=0.9\columnwidth]{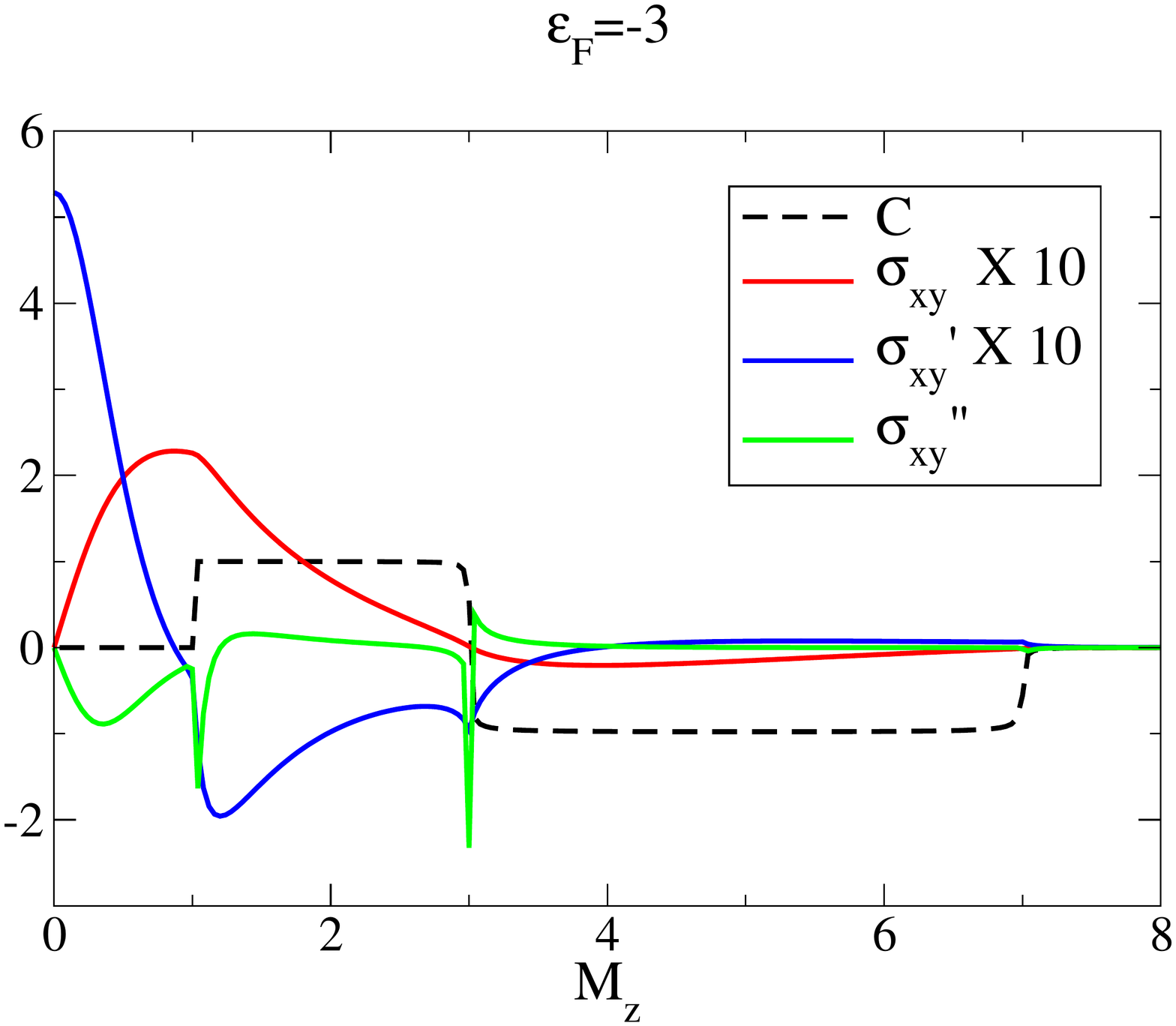}
\includegraphics[width=0.9\columnwidth]{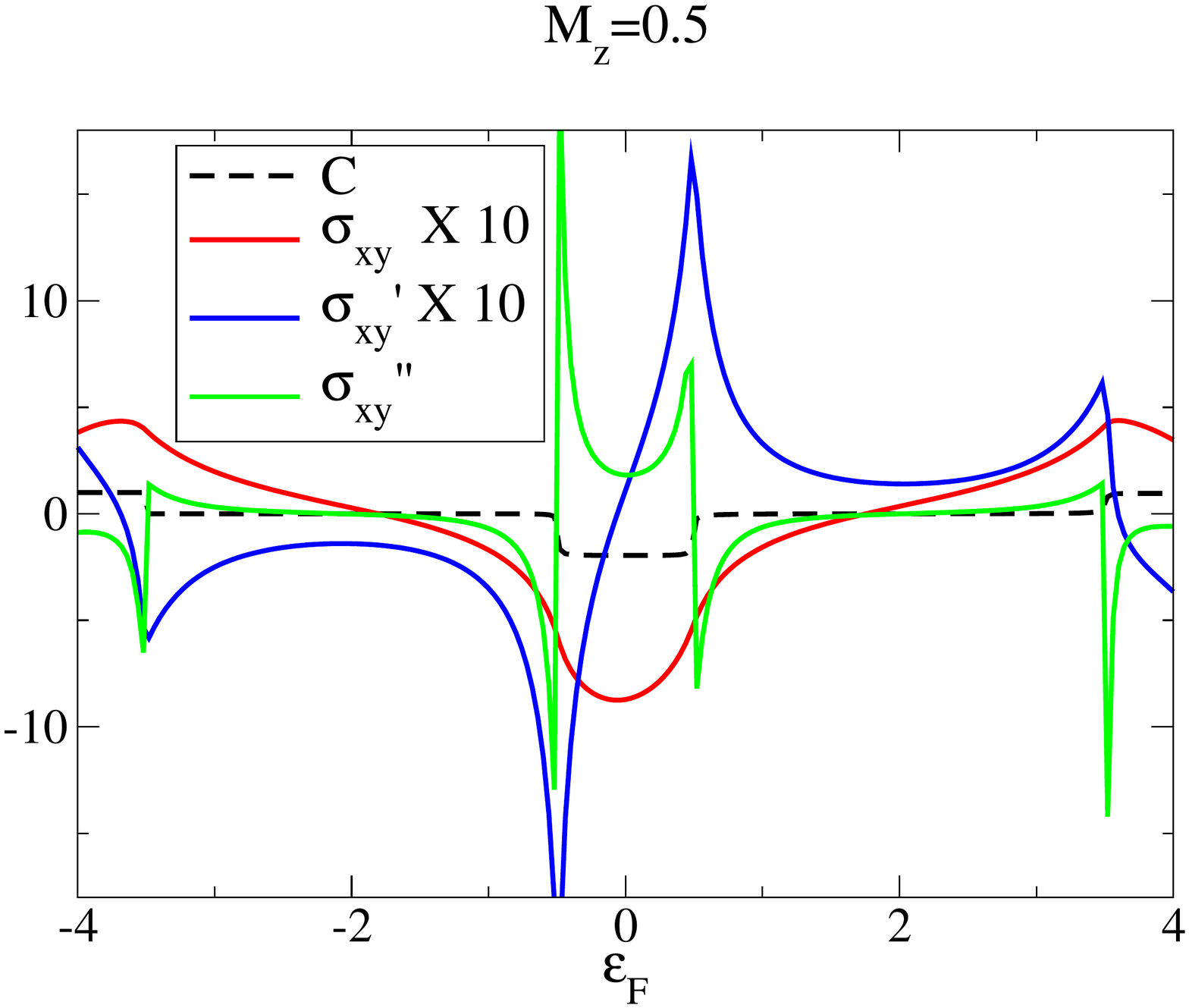}
\par\end{centering}
\caption{\label{fig:hallsder}(color online). Chern number, Hall conductivity and its first and second derivatives 
as a function
of magnetization and as a function of chemical potential for strong spin-orbit coupling.
The Hall conductivity and its first derivative are multiplied by a factor of $10$ for
better visualization.}
\end{figure} 

\begin{figure}
\begin{centering}
\includegraphics[width=0.8\columnwidth]{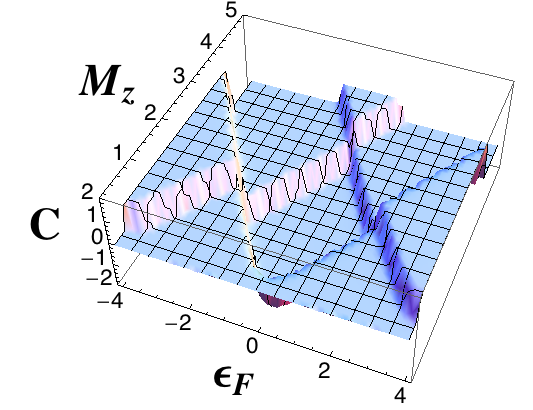}
\includegraphics[width=0.8\columnwidth]{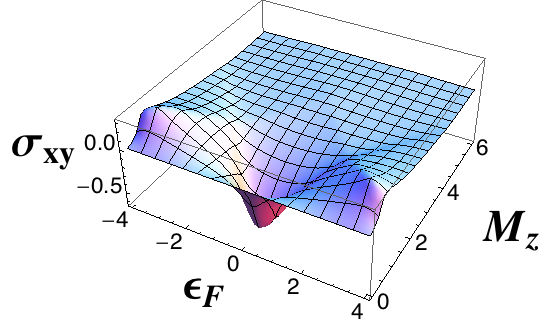}
\par\end{centering}
\caption{\label{fig:hallss}(color online). Chern number and Hall conductivity for s-wave pairing and zero p-wave
pairing ($d=0, \Delta_s=0.5, \alpha=0.6$).
}
\end{figure} 

In the case of a $\mathbb{Z}$~topological insulator, a topological transition modifies the Chern number
and the value of the Hall conductivity 
which, therefore, exhibits a clear signature of the transition.  
In the case of a $\mathbb{Z}$~topological superconductor  there is, however, 
no discontinuity of $\sigma_{xy}$ but its second derivative signals  the transitions sharply.
This is well illustrated in Fig. \ref{fig:hallsder} where we 
show cuts at constant chemical potential
as a function of the magnetization or fixing the magnetization and changing the chemical potential.
The results for the Chern number 
clearly indicate the topological transitions either as
a function of the magnetization or chemical potential. 
The behavior of the Hall conductivity   
correlates with these transitions. 
As expected from general considerations if a transition occurs between Chern
numbers of different signs, the Hall conductivity changes sign accordingly. 
Here we are interested in finding a signature of the change in the  Chern number. This can be achieved by 
looking
at the derivatives of the Hall conductivity. At the transitions the derivative behaves in a way qualitatively
similar to the case of a $\mathbb{Z}$~topological insulator: if the Chern number increases across the transition the 
first derivative of the Hall conductivity is positive  and if the Chern number decreases the derivative is negative. 
The change
is small and the features in the first derivative are also small, therefore, we have multiplied the
Hall conductivity and its first derivative by a factor of $10$. A much stronger
signal is provided by the second derivative.  
At a transition
where the Chern number changes, 
the second derivative exhibits  two close peaks: a negative peak followed by a positive one
when the Chern number increases (and vice-versa if the Chern number decreases).
In the case of 
a topological insulator, the derivative is a Dirac delta
function and the second derivative is the derivative of a Dirac delta function. In the superconductor these
delta functions are smeared but are still 
clear evidence of the location of the transition and, moreover, of the
change in the topological number.   

We may also consider the case where
there is only $s$-wave pairing and no $p$-wave pairing. The results for the Chern number and the Hall
conductivity for the same region of parameters are shown in Fig. \ref{fig:hallss}. 
As shown before, if the magnetization vanishes, and the $s$-wave component is larger than the $p$-wave component,
the phase is topologically trivial.\cite{Sato}
There are non-trivial
phases that arise due to the presence of the magnetization and the consequent breaking of TRS.\cite{SatoPRL09,SauPRL10}
Note that there is a finite region around zero magnetization in which the Chern number vanishes, as mentioned.
As above, the Hall conductivity clearly shows the location of the various transitions between the Chern
numbers and an analysis similar to the one carried out for the $p$-wave case both for the Hall
conductivity and its derivatives may be carried out, signaling in a similar way the various transitions. 
The same holds for weak spin-orbit coupling. The topological transitions are also clearly detected
by the derivatives of the Hall conductivity.

\section{Edge states}  

Due to the bulk-edge correspondence, complementary information on the topological phases and
transitions may be obtained by analyzing the edge states. We consider a strip geometry 
of transversal
width $N_y$ and apply  periodic boundary conditions along the longitudinal direction, $x$.
We write
\be
\psi_{k_x,k_y,\sigma} = \frac{1}{\sqrt{N_y}} \sum_{j_y} e^{-i k_y j_y} \psi_{k_x,j_y,\sigma}\,,
\label{operators}
\ee
and rewrite the Hamiltonian matrix in terms of 
the operators (\ref{operators})  as
\bea
H = \sum_{k_x} \sum_{j_y} 
& & \left(\begin{array}{cccc}
\psi_{k_x,j_y,\uparrow}^{\dagger}  & \psi_{k_x,j_y,\downarrow}^{\dagger} &
\psi_{-k_x,j_y,\uparrow}  & \psi_{-k_x,j_y,\downarrow}
\end{array}\right) \nonumber \\
& & \hat{H}_{k_x,j_y}
\left(\begin{array}{c}
\psi_{k_x,j_y,\uparrow} \\
\psi_{k_x,j_y,\downarrow} \\
\psi_{-k_x,j_y,\uparrow}^{\dagger} \\
\psi_{-k_x,j_y,\downarrow}^{\dagger} \\
\end{array}\right)
\eea

The operator $\hat{H}_{k_x,j_y}$ reads
\begin{widetext}
\be
\left(\begin{array}{cccc}
-2 t \cos k_x -M_z-\epsilon_F-t \eta_+ & i\alpha \sin k_x +\frac{\alpha}{2i} \eta_- &
-i d \sin k_x -\frac{d}{2i} \eta_- & \Delta_s \\
-i \alpha \sin k_x +\frac{\alpha}{2i} \eta_- & -2 t \cos k_x +M_z -\epsilon_F -t \eta_+ &
-\Delta_s & -i d \sin k_x +\frac{d}{2i} \eta_- \\
i d \sin k_x -\frac{d}{2i} \eta_- & -\Delta_s & 
2 t \cos k_x +M_z+\epsilon_F+t \eta_+ & -i\alpha \sin k_x +\frac{\alpha}{2i} \eta_- \\
\Delta_s & i d \sin k_x +\frac{d}{2i} \eta_- &
i \alpha \sin k_x +\frac{\alpha}{2i} \eta_- & 2 t \cos k_x -M_z +\epsilon_F +t \eta_+ \\ 
\end{array}\right)
\ee
\end{widetext}
where $\psi_{j_y}^{\dagger} \eta_{\pm} \psi_{j_y} = \psi_{j_y}^{\dagger} \psi_{j_y+1} \pm \psi_{j_y+1}^{\dagger} \psi_{j_y}$.
The diagonalization of this Hamiltonian involves the solution of a $4 N_y \times 4 N_y$ eigenvalue problem.
The energy states include states in the bulk and states along the edges.

\subsection{Strong spin-orbit coupling}

In the case of strong spin-orbit coupling with $M_z=0$ there is no TRS breaking and the system belongs
to the symmetry class DIII.\cite{LudwigAIP,LudwigNJP} 
In the $s$-wave case there is only the bulk
gap and no gapless (edge) states. The system is in a topologically trivial phase. 
In the case of p-wave
pairing even though the Chern number vanishes there are gapless edge states \cite{Sato}.
The system is in a $\mathbb{Z}_2$ topological phase. The  gapless edge states have a
twofold Kramers degeneracy and 
two counterpropagating edge modes
give opposite contributions to the total Chern number, $C=0$.
This is a similar situation to that 
 in the spin Hall effect, where, even though the charge current vanishes,
there is a spin current along the edges.
In the case where 
there is a mixture of s- and p-wave components and 
the amplitude of the p-wave pairing is larger than the corresponding
amplitude of the s-wave case, there are edge states and a topologically nontrivial phase. 
Because of spin-momentum locking 
there is no backscattering and these states are topologically protected from non-magnetic
impurities.

As the magnetization is turned on TRS is broken
and the system's symmetry class changes to D.\cite{LudwigAIP,LudwigNJP} 
For small magnetization the $\mathbb{Z}$~topological superconductor is in a trivial phase
with Chern number $C=0$. A finite magnetization is then necessary to cause a topological
phase transition to a phase with non-zero Chern number.\cite{Sato}
This happens both for the
p-wave case and the s-wave case. 
The sequence of Chern numbers is clearly correlated with the number of pairs
of edge states as shown in Ref. \onlinecite{Sato}.

\begin{figure}
\begin{centering}
\includegraphics[width=0.8\columnwidth]{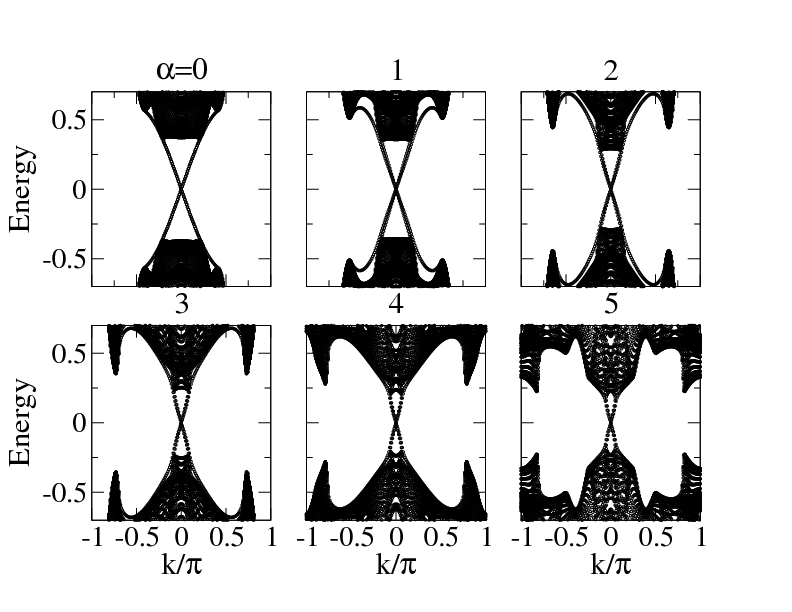}
\par\end{centering}
\caption{\label{fig:edges6}Gapless edge modes for unitary case of strong spin-orbit coupling
for zero Chern number for different values of the spin-orbit coupling, $\alpha$. 
Here $\epsilon_F=-3,~d=0.6,~\Delta=0.1,~M_z=0.5$.
}
\end{figure} 

In Fig. \ref{fig:edges6} we show the edge modes as a function of spin-orbit coupling for the case of
strong spin orbit coupling ($\boldsymbol{d} \parallel \boldsymbol{s}$) 
and small magnetization, when the system is in the $C=0$ phase. The spin orbit coupling does
not change the Chern number since it does not close the bulk gap. 
It is interesting to note that even though the system is in a $C=0$ phase the 
number of edge states is two; the same as that in the parent $\mathbb{Z}_2$ phase, 
when $M_z = 0$.\cite{Sato}
For $M_z \neq 0$, however, TRS is broken and these edge states are not topologically protected against
(any type of) disorder. In this sense the system is in a trivial phase, in accordance with 
the Chern number $C=0$. Nevertheless, in the clean limit these edge modes could be detected.

The presence of edge modes induced by bulk topology can also be shown 
using dimensional reduction and thereby calculating the winding number.\cite{Wen} 
For $k_y=0$ or $\pi$, the Hamiltonian $H(\vk)$ has the chiral symmetry:
\be
\Gamma H(\vk) \Gamma^\dagger = -  H(\vk)
\ee
 with $\Gamma = \tau_x\otimes\sigma_0$, where $\sigma_0$ is the identity in spin space 
and $\tau_x$ acts on the particle-hole space.
The  operator that diagonalizes $\Gamma$ is \cite{Schnyder,Tewari}
\be
T= \sigma_0 \otimes e^{-i\frac{\pi}{4} \tau_y}\,,
\label{eq:T}
\ee
and the Hamiltonian can then be brought to the off-diagonal form:
\be
T H(\vk) T^{\dagger} = 
\left(\begin{array}{cc}
0 & q(\vk) \\
q^{\dagger}(\vk) & 0 
\end{array}\right)\,,
\ee
if $k_y=0,\pi$ and $d_z=0$ where
\bea
q(\vk) = \nonumber \\ 
\left(\begin{array}{cc}
\epsilon_{\vk}-\epsilon_F -M_z+id \sin k_x & i\alpha \sin k_x-\Delta_s \\
-i\alpha \sin k_x+\Delta_s & \epsilon_{\vk}-\epsilon_F +M_z+id \sin k_x 
\end{array}\right)\,. \nonumber \\
\label{eq:qk}
\eea
The winding number is then defined as
\bea
I(k_y) = \nonumber \\ 
\frac{1}{4\pi i} \int_{-\pi}^{\pi} dk_x Tr [ q^{-1}(\vk) \partial_{k_x} q(\vk) -(q^{\dagger})^{-1}(\vk) \partial_{k_x}
q^{\dagger}(\vk) ]\,. \nonumber \\
(k_y=0,\pi) \nonumber\\
\eea
Physically, a nonzero $I(k_y)$ means that if the system is infinite along the $y$ direction and finite along $x$,
there will be edge states with $k_y=0$ or $\pi$.\cite{yakovenko}
The calculation of the winding number gives the number of gapless edge modes both when the Chern number vanishes
and when the Chern number is finite.\cite{Sato}

\subsection{Weak spin-orbit coupling}

\begin{figure}
\begin{centering}
\includegraphics[width=0.8\columnwidth]{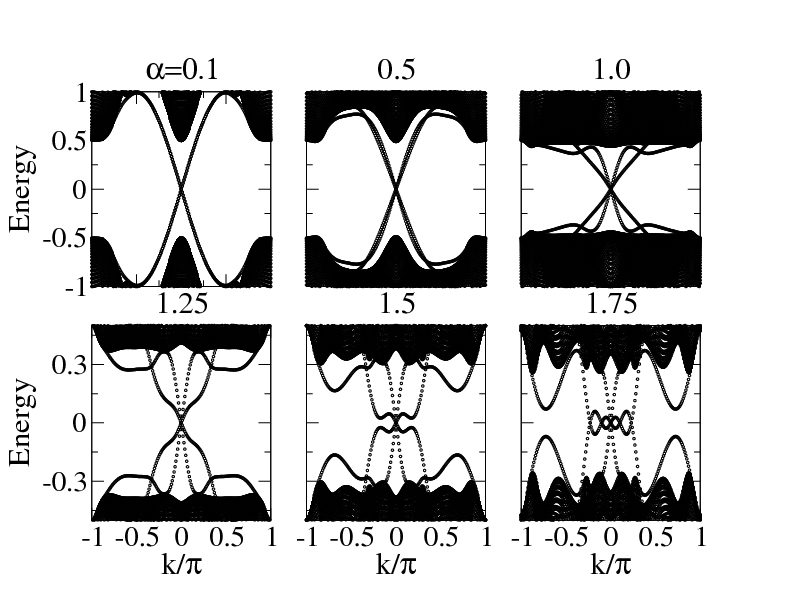}
\includegraphics[width=0.8\columnwidth]{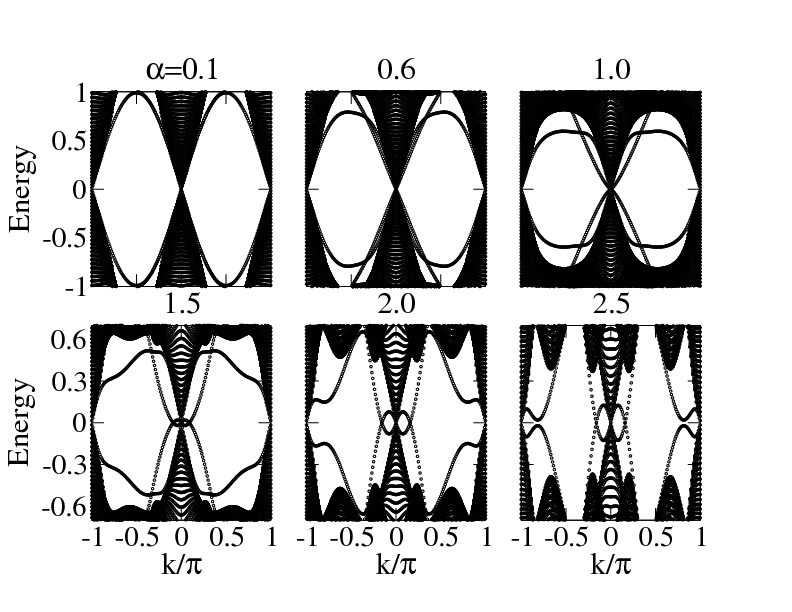}
\par\end{centering}
\caption{\label{fig:edges7}Gapless edge modes in the unitary case for 
zero Chern number (top) and Chern number equal to $2$ (bottom). 
Here $\epsilon_F=-1,d=1$ and $M_z=0.5$ (top), $M_z=1.2$ (bottom).}
\end{figure} 

If the spin-orbit coupling is not strong, so that the pairing vector $\boldsymbol d$ is not aligned with the spin-orbit
vector, as in the unitary and non-unitary cases considered in Sec.~\ref{sec:Csigma}, 
the connection between the Chern number and the number of gapless edge states is less
transparent.  Fig.~\ref{fig:edges7} shows the low-lying energy modes for the unitary case previously
considered for different values of the spin orbit coupling, $\alpha$. The top panel corresponds to a 
case where the Chern number vanishes while the bottom panel
to a non-vanishing Chern number. There is a variety of gapless edge states 
 for both the $C=0$ and $C=2$ cases. 
Even though changing $\alpha$ should not change the topology, there is an apparent appearance of various
gapless states that seem not to  follow the bulk-edge correspondence. However, 
in the unitary case Eq.~\eqref{eq:T} still transforms 
the Hamiltonian into an off-diagonal form similar to that in Eq.~\eqref{eq:qk},
so that the winding number is still well defined.
A calculation of the winding
number shows that the number of gapless edge modes is actually independent of $\alpha$.
In the top panel
($C=0$) we get that $I(0)=2$ and $I(\pi)=0$ and in the bottom panel we obtain that $I(0)=1$ and $I(\pi)=1$
in agreement with the value $C=2$.

\begin{figure}
\begin{centering}
\includegraphics[width=0.9\columnwidth]{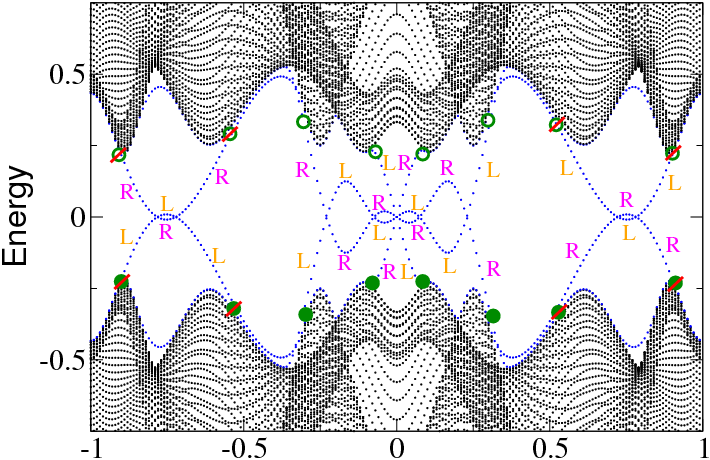}
\includegraphics[width=0.9\columnwidth]{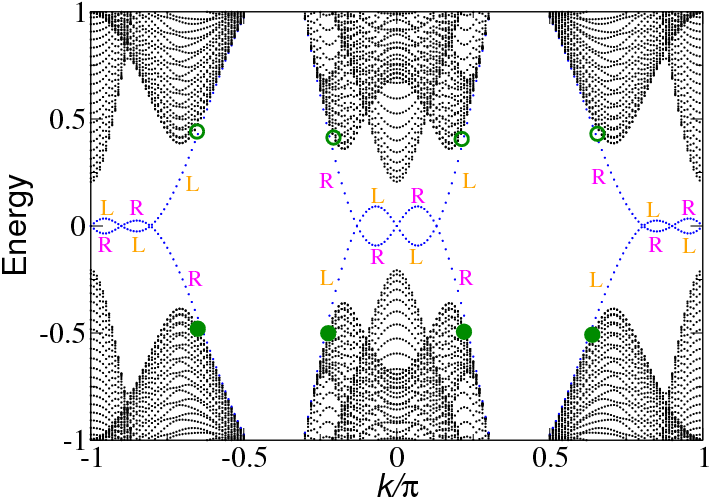}
\par\end{centering}
\caption{\label{fig:edges8}(color online). Gapless edge modes for unitary case for zero Chern number (top) and
Chern number equal to $2$ (bottom). 
Here $\epsilon_F=-1,d=1$ and $M_z=0.5,\alpha=2$ (top), and $M_z=1.2,\alpha=3$ (bottom).}
\end{figure} 

The bulk-edge correspondence is further elucidated in Fig.~\ref{fig:edges8}. 
Careful analysis shows that some of the
gapless edge states do not originate
from bulk topology and that the number of topologically 
induced edge states (given either by
the winding number or the Chern number in the case of $C\neq 0$) is consistent. Only the bands of edge states that connect
the upper and lower bulk bands, i.e. connecting open and filled circles in Fig.~\ref{fig:edges8}, 
can be traced back to the nontrivial bulk topology.\cite{Hatsugai2} Denoting the two edges of the system as $R$ and $L$
we see that 
 for $C=0$, the number of propagating states at each edge is always the same as the number of counter propagating ones.
For $C=2$, on the other hand, the difference between propagating and counter propagating states is always~2 at each edge.

\begin{figure}
\begin{centering}
\includegraphics[width=0.8\columnwidth]{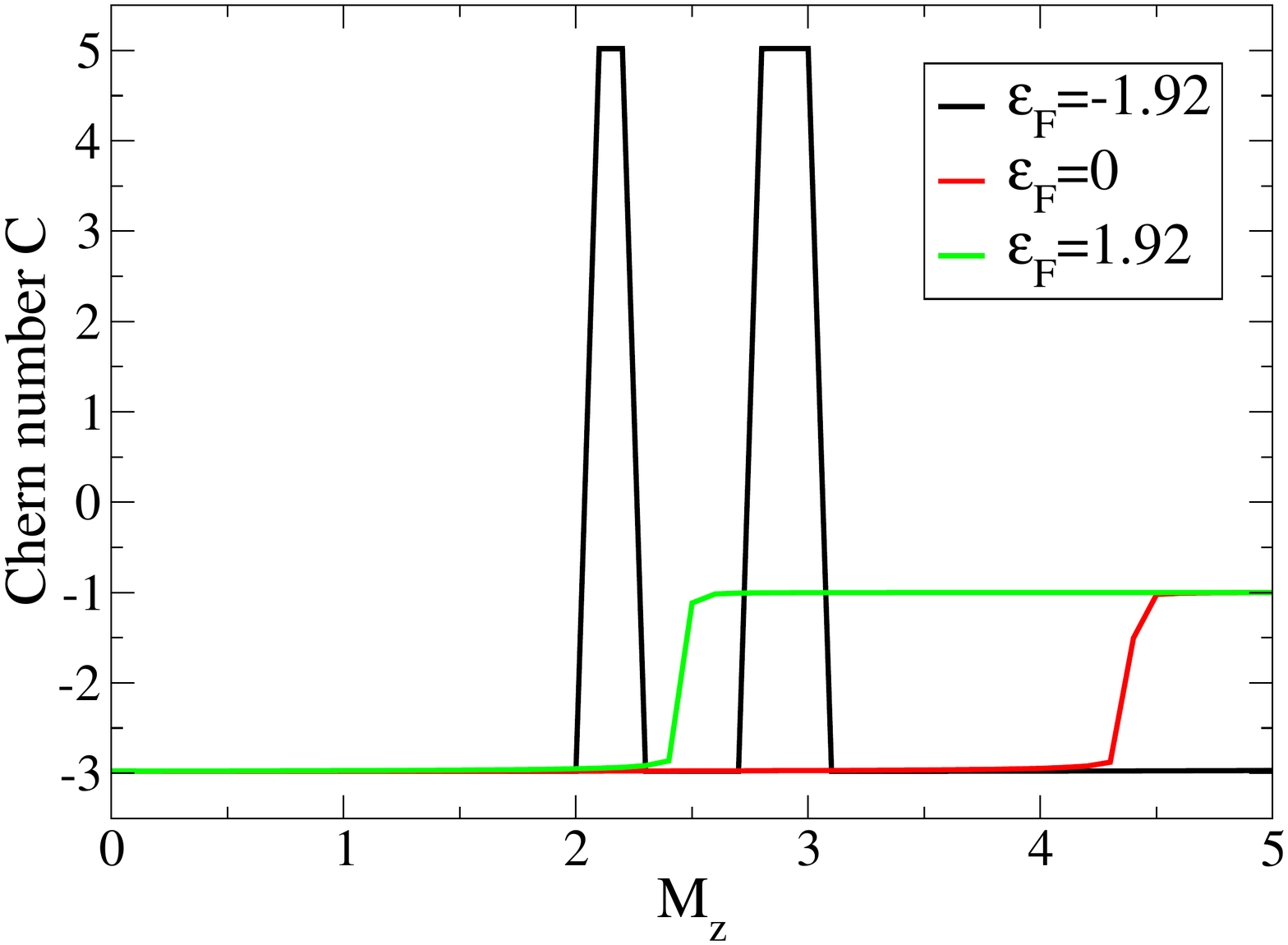}
\includegraphics[width=0.8\columnwidth]{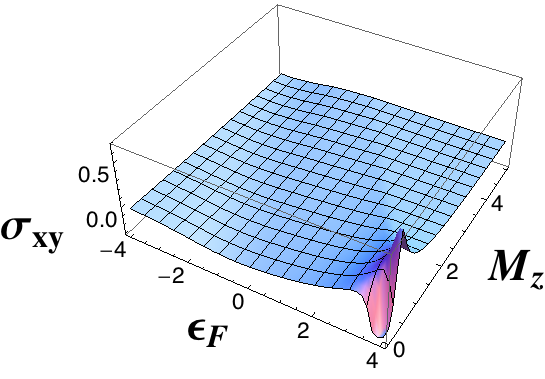}
\par\end{centering}
\caption{\label{fig:nont}(color online). Chern number and Hall conductivity 
in the case where the normal system is topologically nontrivial. 
The parameters are: $t_1=1,t_2=1.1,\Delta_s=0.1,d=0.6,\alpha=0.6$.
In the lower panel $\epsilon_F=1.92$. }
\end{figure} 
  
\section{Nontrivial topology in normal phase}

The nontrivial topology of the bands in the superconducting phases above
stems from the mixture of the particle and hole
excitations since the normal phase is non-topological.
We may as well
consider a system that is nontrivial in the normal phase and add superconductivity either self-consistently
or via a proximity effect. This has been proposed before in various contexts \cite{Fu}. 

\begin{figure}
\begin{centering}
\includegraphics[width=0.8\columnwidth]{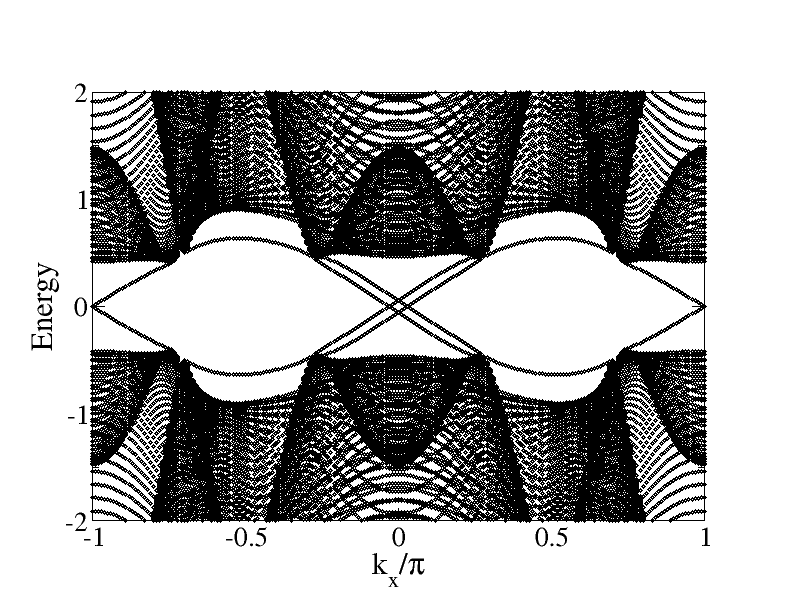}
\includegraphics[width=0.8\columnwidth]{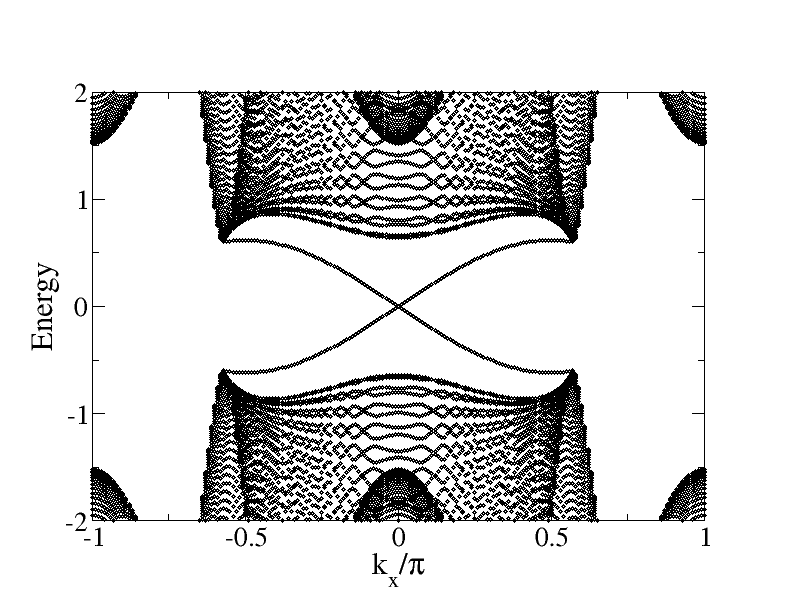}
\par\end{centering}
\caption{\label{fig:nont2}Gapless edge modes for 
the same system as in Figure \ref{fig:nont} with $\epsilon_F=1.92$ 
  and:   $M_z=1$, $C=-3$  (top);
 $M_z=4$,  $C=-1$  (bottom).
}
\end{figure} 

Considering a Hamiltonian of the form of Eq. \eqref{H} and selecting 
\begin{eqnarray}
h_x &=& \alpha\sin k_y\,, \qquad h_y = -\alpha\sin k_x \,,  \nonumber\\  
h_z &=& 2t_1 \left( \cos k_x + \cos k_y \right) + 4 t_2 \cos k_x \cos k_y\,,
\end{eqnarray}
leads to nontrivial phases as the hoppings $t_1$ and $t_2$ are varied \cite{beijing}. 
Results for the edge states, the Chern number and the Hall conductivity are shown in Figs. \ref{fig:nont} and
\ref{fig:nont2}.
In the normal phase TRS
 is broken since $h_z$ is even in the momentum. The
Chern number 
is  
$C= 2$ if $| t _1| < |t_2 |$;
and    $C= 1$ if  $| t_1 | > |t_2  |$.
As the system becomes superconducting the
 nontrivial topology remains even though the Chern number changes. 
The nontrivial topology of the normal
state bands lends some robustness to the topological superconducting phase. 
Indeed, the Chern number remains invariant in large portions of the parameter space. 
In the first panel of Fig. \ref{fig:nont} we show cuts of the Chern number at constant chemical potential,
$\epsilon_F=-1.92,0,1.92$, as a function of $M_z$. For negative chemical potential the Chern number is $C=-3$
except for some narrow regions where $C=5$. For zero and positive chemical potential there is a single
topological transition from $C=-3$ to $C=-1$.
The results for the edge states in Fig.~\ref{fig:nont2} 
show a clear correspondence between the Chern number and the number of gapless modes. 
Note that in this case there is strong spin-orbit coupling.
The difference to the previous Sections is the 
non-trivial topology of the normal phase.
The superconducting pairing then changes the topology,
as shown by the change of Chern number entering the superconducting phase.
In this case the Hall conductivity 
also signals the transition at positive values of the chemical potential 
and varies smoothly in
the narrow region where $C=5$.  

\section{Conclusions}
  
We have shown that the Hall conductivity and its derivatives may be used to detect
the topological transitions that occur in $\mathbb{Z}$~topological superconductors. In a topological insulator the Hall conductivity
is quantized and proportional to the Chern number and, therefore, its discontinuous changes across a transition
can be used to detect and characterize the transition. Even though the Hall conductivity is not quantized in 
a superconductor \cite{pdss1999} it may also be used to study these transitions. This provides a bulk detection method of these
transitions that is complementary to the detection of the gapless edge states associated with these nontrivial
topological phases.

In the case of strong spin-orbit coupling there is a simple correspondence between the number of gapless
edge states and the Chern number, both for trivial and nontrivial normal state bands.
However, in the case of weak spin orbit coupling where the pairing vector, $\boldsymbol{d}$, is not parallel
to the spin-orbit vector, $\boldsymbol{s}$, 
extra unprotected gapless modes appear. The bulk-edge correspondence is preserved as evidenced by
the calculation of the winding number.


\begin{thebibliography}{99}
\bibitem{errata}
Note that in Ref. \cite{us} the amplitude $v_n(-\boldsymbol{k},\uparrow)$ in Eq. (2) is not affected by
a minus sign. This sign change leads to numerical results for the Hall conductivity presented
in that reference that are not correct and are corrected in this manuscript. 
The conclusions remain valid as there is no qualitative
change of the results.
\end{thebibliography}
\end{document}